\documentclass{aa}  

\usepackage{graphicx}
\usepackage{txfonts}
\usepackage{graphicx}   % Including figure files
\usepackage{amsmath}    % Advanced maths commands
\usepackage{amssymb}    % Extra maths symbols
\usepackage{color}      % text color
\usepackage{natbib}
\usepackage{subcaption}
\usepackage{footnote}
\usepackage{threeparttable}
\makesavenoteenv{tabular}
\makesavenoteenv{table}

\bibpunct{(}{)}{;}{a}{}{,} % to follow the A&A style
\newcommand\newsubcap[1]{\phantomcaption%
       \caption*{\figurename~\thefigure\thesubfigure: #1}}

\newcommand\starname{NGTS-5}
\newcommand\planetname{NGTS-5b}
\newcommand{\PrimaryTeff}{$4987  \pm 41$}

\newcommand{\PrimaryLogglc}{$4.52  \pm 0.037$}
\newcommand{\PrimaryMet}{$0.12   \pm 0.1$}
\newcommand{\PrimaryMass}{$0.661 \substack{+0.068 \\ -0.061}$}
\newcommand{\PrimaryRadius}{$0.739 \substack{+0.014 \\ -0.012}$}

\newcommand\ms{\hbox{m\,s$^{-1}$}}  %m per sec
\newcommand\teff{$T_{\rm{eff}}$}
\newcommand\logg{{\it log}\,{\it g$_\ast$}}
   
\newcommand\kms{\hbox{km\,s$^{-1}$}}  %km per sec
\newcommand\Msun{\hbox{$M_{\odot}$}}  %Msun
\newcommand\Rsun{\hbox{$R_{\odot}$}}  %Rsun
\newcommand\Mjup{\hbox{$M_\mathrm{Jup}$}}  %Mjup
\newcommand\Rjup{\hbox{$R_\mathrm{Jup}$}}  %Rjup

\newcommand\loggp{{\it log}\,{\it g$_P$}}
\newcommand\Mstar{\hbox{$M_{*}$}}  %Msun
\newcommand\Rstar{\hbox{$R_{*}$}}  %Rsun

\newcommand{\PERIOD}{$3.3569866 \pm 0.0000026$}
\newcommand{\EPOCH}{$7740.35262 \pm 0.00026$}
\newcommand{\ECCENTRICITY}{$0.0 \mbox{ }(fixed)$}
\newcommand{\DURATION}{$2.21 \pm 0.03$}

\newcommand{\KAmp}{$41.0 \substack{+5.9 \\ -6.4}$}
\newcommand{\GAMMA}{$-5.5225 \substack{+0.0039 \\ -0.0042}$}
\newcommand{\SSMA}{$11.111 \substack{+0.315 \\ -0.296}$}
\newcommand{\SMA}{$0.0382 \pm 0.0013$}
\newcommand{\SSR}{$0.1579 \pm 0.0016$}

\newcommand{\INCLINATION}{$86.6 \pm 0.2$}

\newcommand{\LDOnez}{$0.29 \pm 0.02$}
\newcommand{\LDTwoz}{$0.40 \pm 0.02$}
\newcommand{\LDOnei}{$0.35 \pm 0.03$}
\newcommand{\LDTwoi}{$0.41 \pm 0.02$}
\newcommand{\LDOnev}{$0.59 \pm 0.04$}
\newcommand{\LDTwov}{$0.48 \pm 0.02$}
\newcommand{\LDOnen}{$0.44 \pm 0.03$}
\newcommand{\LDTwon}{$0.44 \pm 0.02$}

\newcommand{\SecondaryTeff}{$952 \pm 24$}
\newcommand{\SecondaryMass}{$0.229 \pm 0.037$}
\newcommand{\SecondaryRadius}{$1.136 \pm 0.023$}
\newcommand{\SecondaryRadiuspm}{$1.136 \pm 0.023$}
\newcommand{\SecondaryLogg}{$2.643 \substack{+0.066 \\ -0.078}$}

\newcommand{\noprog}{098.C-0820(A), 099.C-0303(A), 0100.C-0474(A), and 0101.C-0623(A)}

\newcommand{\gpe}{GP--EBOP}

\begin{document}

   \title{\planetname:{}  a highly inflated planet offering insights into the sub-Jovian desert\thanks{
   Based on observations made with CORALIE echelle spectrograph mounted on the 1.2~m Swiss telescope and the HARPS spectrograph on the ESO 3.6~m telescope, both at La Silla observatory (ESO, Chile). HARPS programme IDs are \noprog}}

   \author{Philipp Eigm\"uller
          \inst{1,2}\fnmsep\thanks{\email{philipp.eigmueller@dlr.de}}
          \and Alexander Chaushev\inst{2}
          \and Edward~Gillen\inst{7,11}
          \and Alexis Smith\inst{1}
          \and Louise~D.~Nielsen \inst{8}
          \and Oliver Turner\inst{8}
          \and Szilard Czismadia\inst{1}
          \and Barry Smalley\inst{16}
          \and Daniel~Bayliss \inst{4,5}
          \and Claudia Belardi \inst{6}
          \and Fran\c{c}ois Bouchy \inst{8}
          \and Matthew R. Burleigh\inst{6}
          \and Juan Cabrera\inst{1}
          \and Sarah L. Casewell\inst{6}
          \and Bruno Chazelas\inst{8}
                  \and Benjamin F. Cooke\inst{4,5} 
                  \and Anders Erikson\inst{1}
          \and Boris~T.~G\"ansicke\inst{4,5}
          \and Maximilian~N.~G{\"u}nther\inst{7}
          \and Michael R.~Goad\inst{6}
          \and Andrew Grange\inst{6}
          \and James~A.~G.~Jackman\inst{4,5}
          \and James~S.~Jenkins\inst{9,10}
          \and James McCormac\inst{4,5}
          \and Maximiliano Moyano\inst{13}
          \and Don~Pollacco\inst{4,5}
          \and Katja Poppenhaeger\inst{12,14,15}
          \and Didier~Queloz\inst{7}
          \and Liam~Raynard\inst{6}
          \and Heike Rauer\inst{1,2,3}
          \and St\'{e}phane~Udry\inst{8}
          \and Simon.~R.~Walker\inst{4,5}
          \and Christopher~A.~Watson \inst{12}
          \and Richard~G.~West\inst{4,5}
          \and Peter~J.~Wheatley\inst{4,5}
          }

   \institute{Institute of Planetary Research, German Aerospace Center, 
              Rutherfordstrasse 2, 12489 Berlin, Germany              
         \and Center for Astronomy and Astrophysics, TU Berlin, Hardenbergstr. 36, D-10623 Berlin, Germany
         \and Institute of Geological Sciences, FU Berlin, Malteserstr. 74-100, D-12249 Berlin, Germany
         \and Centre for Exoplanets and Habitability, University of Warwick, Gibbet Hill Road, Coventry CV4 7AL, UK
         \and Dept.\ of Physics, University of Warwick, Gibbet Hill Road, Coventry CV4 7AL, UK
         \and Department of Physics and Astronomy, Leicester Institute of Space and Earth Observation, University of Leicester, LE1 7RH, UK
         \and Astrophysics Group, Cavendish Laboratory, J.J. Thomson Avenue, Cambridge CB3 0HE, UK
         \and Observatoire de Gen{\`e}ve, Universit{\'e} de Gen{\`e}ve, 51 Ch. des Maillettes, 1290 Sauverny, Switzerland
         \and Departamento de Astronomia, Universidad de Chile, Casilla 36-D, Santiago, Chile
         \and Centro de Astrof\'isica y Tecnolog\'ias Afines (CATA), Casilla 36-D, Santiago, Chile
         \and Astrophysics Research Centre, School of Mathematics and Physics, Queen's University Belfast, BT7 1NN Belfast, UK
         \and Winton Fellow
         \and Instituto de Astronomía, Universidad Católica del Norte, Angamos 0610, 1270709, Antofagasta, Chile
         \and Leibniz Institute for Astrophysics Potsdam (AIP), An der Sternwarte 16, D-14482 Potsdam, Germany
\and Institute for Physics and Astronomy, University of Potsdam, Campus Golm, Karl-Liebknecht-Str. 24/25, D-14476 Potsdam, Germany
\and Astrophysics Group, Keele University, Staffordshire ST5 5BG, UK
}

   \date{Received September 15, 1996; accepted March 16, 1997}

% \abstract{}{}{}{}{} 
% 5 {} token are mandatory
 
  \abstract
   {Planetary population analysis gives us insight into formation and evolution processes. For short-period planets, the sub-Jovian desert has been discussed in recent years with regard to the planet population in the mass/period and radius/period parameter space without taking stellar parameters into account. The Next Generation Transit Survey (NGTS) is optimised for detecting planets in this regime, which allows for further analysis of the sub-Jovian desert.}
   {With high-precision photometric surveys (e.g. with NGTS and TESS), which aim to detect short period planets  especially  around M/K-type host stars, stellar parameters need to be accounted for when  empirical data are compared to model predictions. Presenting a newly discovered planet at the boundary of the sub-Jovian desert, we analyse its bulk properties and use it to show the properties of exoplanets that border the sub-Jovian desert.}
   {Using NGTS light curve and spectroscopic follow-up observations, we confirm the planetary nature of planet \planetname{} and determine its mass. Using exoplanet archives, we set the planet in context with other discoveries.}
   {\planetname{} is a short-period planet with an orbital period of \PERIOD{} days. With a mass of \SecondaryMass \, \Mjup \, and a radius of \SecondaryRadiuspm  \, \Rjup, it is highly inflated. Its mass places it at the upper boundary of the sub-Jovian desert. Because the host is a K2 dwarf, we need to account for the stellar parameters when \planetname{} is analysed with regard to planet populations.}
  {With red-sensitive surveys (e.g. with NGTS and TESS), we expect many more planets around late-type stars to be detected. An empirical analysis of the sub-Jovian desert should therefore take stellar parameters into account.}
   %{With red-sensitive surveys (e.g. NGTS, TESS), we expect much more detections of planets around late type stars to be detected. The empirical analysis of the sub-Jovian desert thus should take the stellar parameters into account.}
  % {With red-sensitive surveys (e.g. NGTS, TESS), we expect much more detections of planets around late type stars to be detected. To account for the different stellar types of planetary host stars, the sub-Jovian desert should not be analysed in the mass period or radius/period plane.}

   \keywords{Planetary systems; Planets and satellites: detection; Planets and satellites: gaseous planets}

\maketitle
%
%________________________________________________________________
\section{Introduction}
Thousands of extrasolar planets have been discovered by now, several hundred of which are well characterised by their bulk properties. Although these exoplanets are distributed over a wide parameter space, it is evident that we have reached a time when we can perform detailed population analysis and set our discoveries in context of planet formation and planetary system evolution theories. 
One of the discovered groups are the very hot short-period planets with orbital periods of only a few days.
Population analysis of these short-period planets that were discovered with ground-based and space-borne surveys suggests a significant dearth of hot-Neptunian planets. This so called "sub-Jovian desert" cannot be explained by observational bias but seems to be related to the physical processes in the evolution of these hot planets. A detailed empirical study of the location of the sub-Jovian desert has been carried out by \citet{Mazeh2016}. %They analyze the planet population for these short period planets in the mass-period as well as in the radius-period parameter space. 
They analysed the population for these short-period planets in the mass-period and radius-period parameter space. 
Although the absence of short-period planets for a certain mass or radius has been proven, it is not clear which physical processes are responsible.\\
Possible effects include photo-evaporation \citep{Lundkvist2016}, which strips the planet of its atmosphere and reduces it in size and mass. This could explain especially the lower boundary of the sub-Jovian desert in relation to planetary mass as well as in regard to radius. Photo-evaporation is related to the incidence of  UV/X-ray photons that heat the upper atmosphere of the planet. Another observed effect is the inflation of hot Jupiters, a phenomenon that is related to the equilibrium temperature of the planet. It is not completely understood what causes this inflation. Possible explanations amongst others include ohmic heating \citep{Laughlin2011}, tidal dissipation \citep{Bodenheimer2001}, kinetic heating \citep{Guillot2002}, and enhanced atmospheric opacities \citep{Burrows2007}. The recent study of inflated radii of hot-Jupiter planets by \citet{Sestovic2018} analysed the relationship of radius to stellar irradiation and mass. Radius inflation might play an important role in the location and shape of the upper boundary of the sub-Jovian desert in the radius/period parameter space. Independent of these effects, \citet{Matsakos2016} showed that high-eccentricity migration alone can explain the observed population. A recent study by \citet{Owen2018} combined the effects of photo-evaporation, planetary inflation, and high-eccentricity migration to explain the observed planet distribution in the mass/period and radius/period parameter space.\\
Meanwhile, the number of detected short-period planets increases with highly sensitive ongoing and new surveys (e.g. K2 \citep{Howell2014}, Next Generation Transit Survey (NGTS) \citep{Wheatley2018}, TESS \citep{Ricker2010}). Some of these discoveries lie at the very edge or even within the sub-Jovian desert \citep[e.g.][]{Eigmuller2017,West2018}. 
With the ongoing high-sensitivity surveys that target short-period planet discoveries, we expect to find many more planets in this regime. As these new surveys are optimised towards redder stars, most of these planets will likely be detected around smaller stars. 
This will allow us to analyse the sub-Jovian desert not only as a homogeneous sample, but will also permit us to investigate the planet distribution as a function of different stellar parameters such as mass, luminosity, temperature, and metallicity, presumably allowing tests of prediction by different explanations for the desert.\\
Presenting a short-period planet around a K-type star that was newly discovered by NGTS, we discuss the sub-Jovian desert and how the stellar type could be accounted for in future discussions of the dearth of short-period sub-Jovian planets. First we concentrate on the newly discovered planet \planetname. In section  \ref{sec::obs} the observations are described, followed by a description of the analysis (Sec. \ref{sec::ana}) and the resulting parameters for \planetname{} (Sec. \ref{sec::res}). In section \ref{sec::dis} we  discuss \planetname{} and place it in context of other planets. We also discuss the effect of the host star on the sub-Jovian desert. Finally, we summarise our results in section \ref{sec::con}.

%__________________________________________________________________

\section{Observations}\label{sec::obs}
\subsection{NGTS observation}
NGTS \citep{2012SPIE.8444E..0EC,Wheatley2018} is a dedicated ground-based transit survey that photometrically monitors millions of stars between 10th and 15th magnitude in V\,mag. It consists of 12 independent telescopes each with a field of view of 8 square degrees. The red-sensitive CCD cameras allow us to detect signals of 1\,mmag for stars brighter than 13th magnitude. The main goal of the survey is to detect Neptune-sized planets around K-type stars. To allow us to detect planets around low-mass stars, the photometric bandwidth of the instrument was selected to be 580-920nm. A search for planets around low-mass stars offers two advantages. 
On the one hand, the number of confirmed planets around M- and K-type stars is still quite low and is the limiting factor of statistical analyses of planetary populations around host stars of different stellar types. On the other hand, a planet around a small star produces a much stronger transit and radial velocity signal than a planet around a G-type star.\\
Regular observations of NGTS started in April of 2016, and the first discoveries \citep[e.g.][]{Bayliss2018,Raynard2018,Gunther2018,West2018,Jackman2018} have recently been reported. 
The first NGTS data release is publicly available in the ESO archive.\\
For the transit survey, each telescope is pointed independently at a single field. \starname{} has been observed from January until September 2016 on 112 observing nights. All together over 85,000 observations, each with 10s exposure time, have been gathered by NGTS.\\

\begin{figure}
        \centering
        \includegraphics[width=0.95\columnwidth]{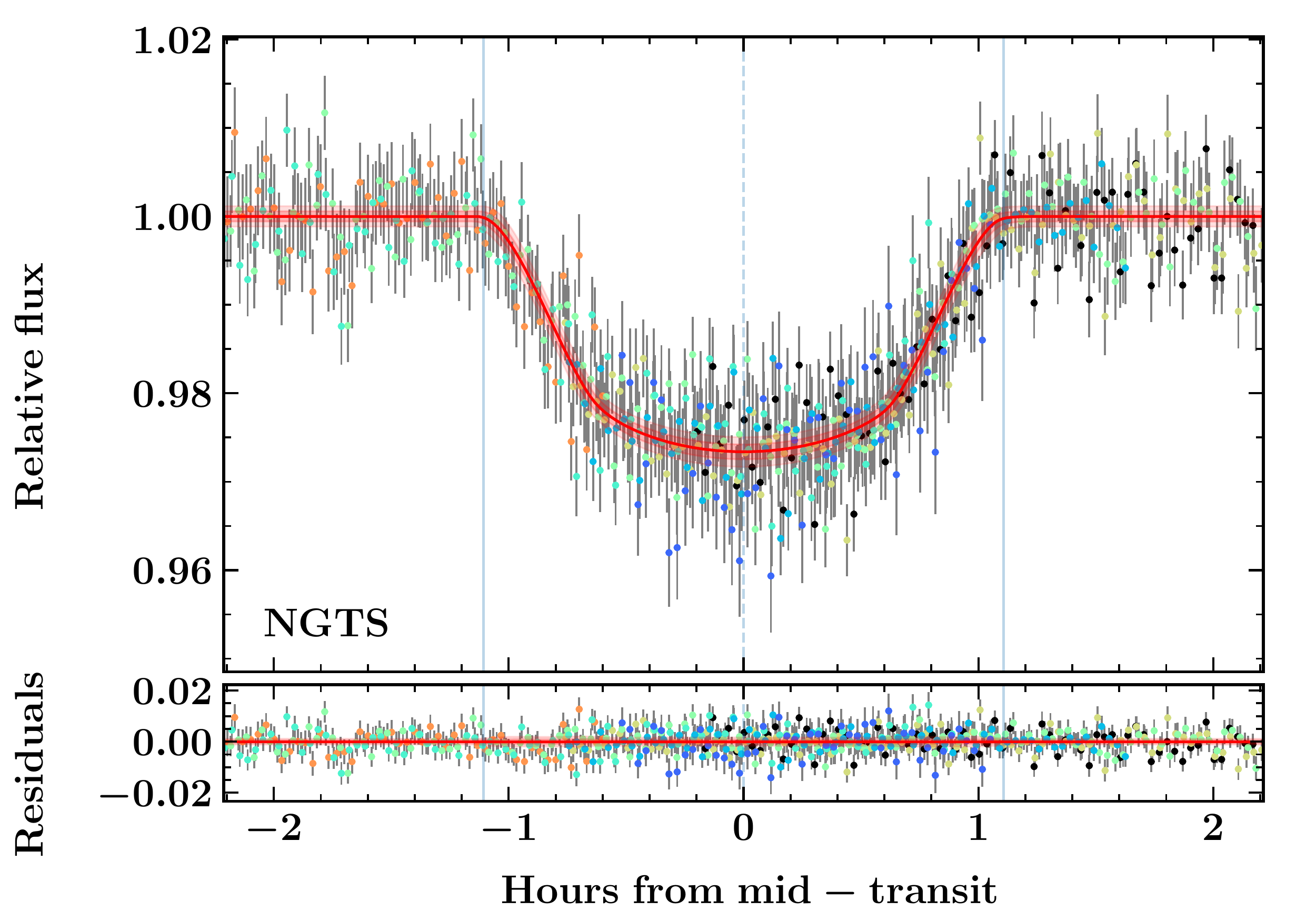}
        \caption{Phase-folded light curve including NGTS data. The red line and pink shaded region show the median and 2\,$\sigma$ confidence intervals of the posterior GP-EBOP model \citep{Gillen2017}. Differently coloured points represent data from different transits.}
    \label{fig::ngtsphase}
\end{figure}

The NGTS data were processed using the NGTS pipeline as described in \citet{Wheatley2018}. 
First a standard data reduction, including a bias subtraction and a flat field correction, was applied to all frames. 
To generate the light curve, aperture photometry based on the CASUtools \citep{2004SPIE.5493..411I} software package was performed. To remove the most dominant systematic effects, the SysRem algorithm \citep{2005MNRAS.356.1466T} was used. Finally, using our own implementation of the BLS algorithm \citep{2002A&A...391..369K}, \planetname{} was detected. The phase-folded light curve is shown in Figure \ref{fig::ngtsphase}. When we vetted the candidate, we found no signs that would indicate an eclipsing binary system. We also used the centroid vetting procedure \citep{guenther2017} to rule out contamination from background objects. After passing all vetting steps, we followed-up \planetname{} to confirm the planetary nature of the system and measure the planetary parameters.

\subsection{Photometric follow-up}
\begin{figure}
        \centering
        \rotatebox{00}{\tiny SAAO (V) \hspace{2.8cm} Euler (I)}\\
        \rotatebox{90}{\tiny Residuals  Relative flux}
    \includegraphics[trim= 50 50 0 40,clip,width=0.45\columnwidth]{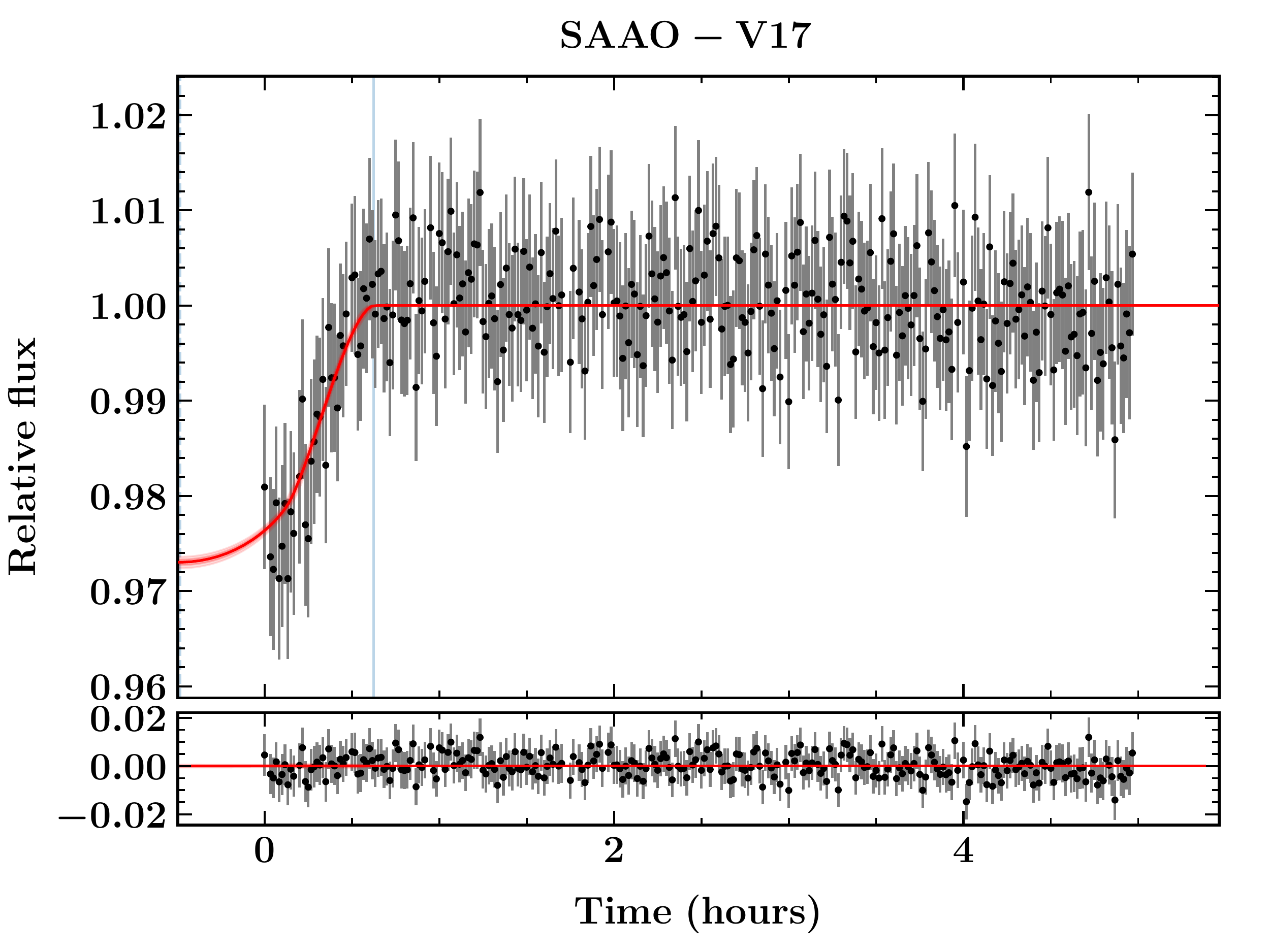}
    \includegraphics[trim= 50 50 0 40,clip,width=0.45\columnwidth]{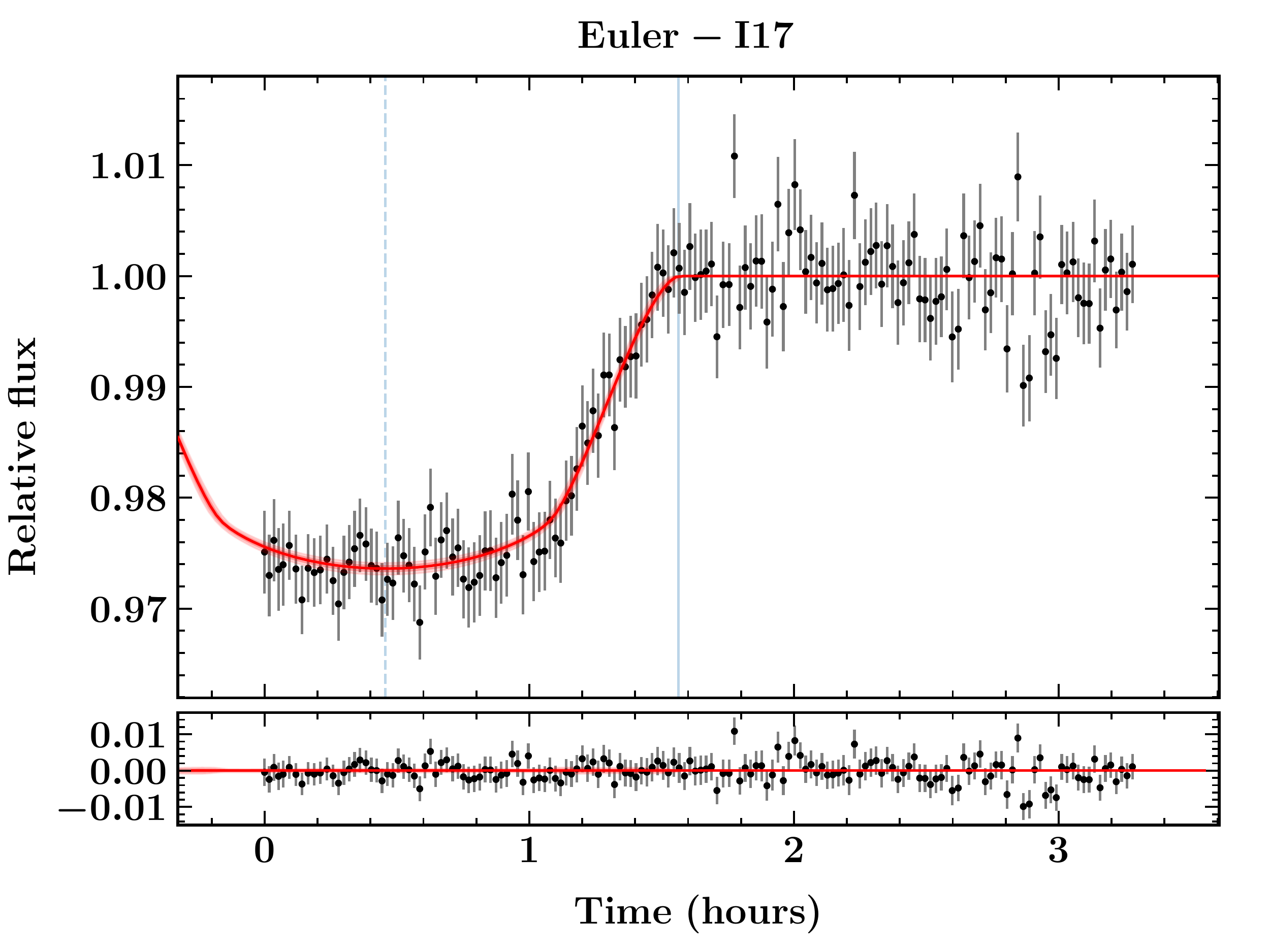}\\
        \rotatebox{00}{\tiny Euler (Z) \hspace{2.8cm} Euler (Z)}\\
        \rotatebox{90}{\tiny Residuals \mbox{      } Relative flux}
    \includegraphics[trim= 50 50 0 40,clip,width=0.45\columnwidth]{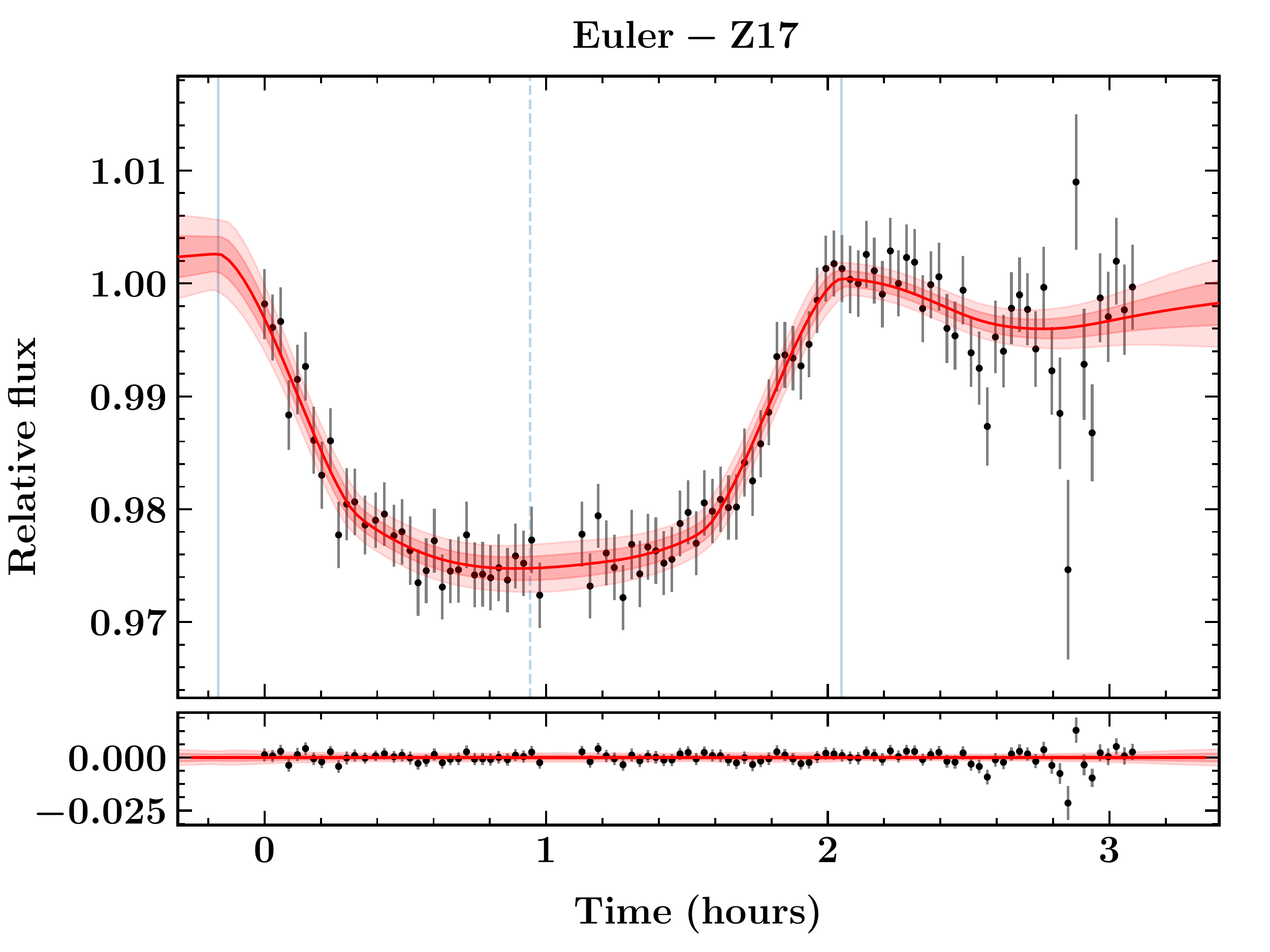}
    \includegraphics[trim= 50 50 0 40,clip,width=0.45\columnwidth]{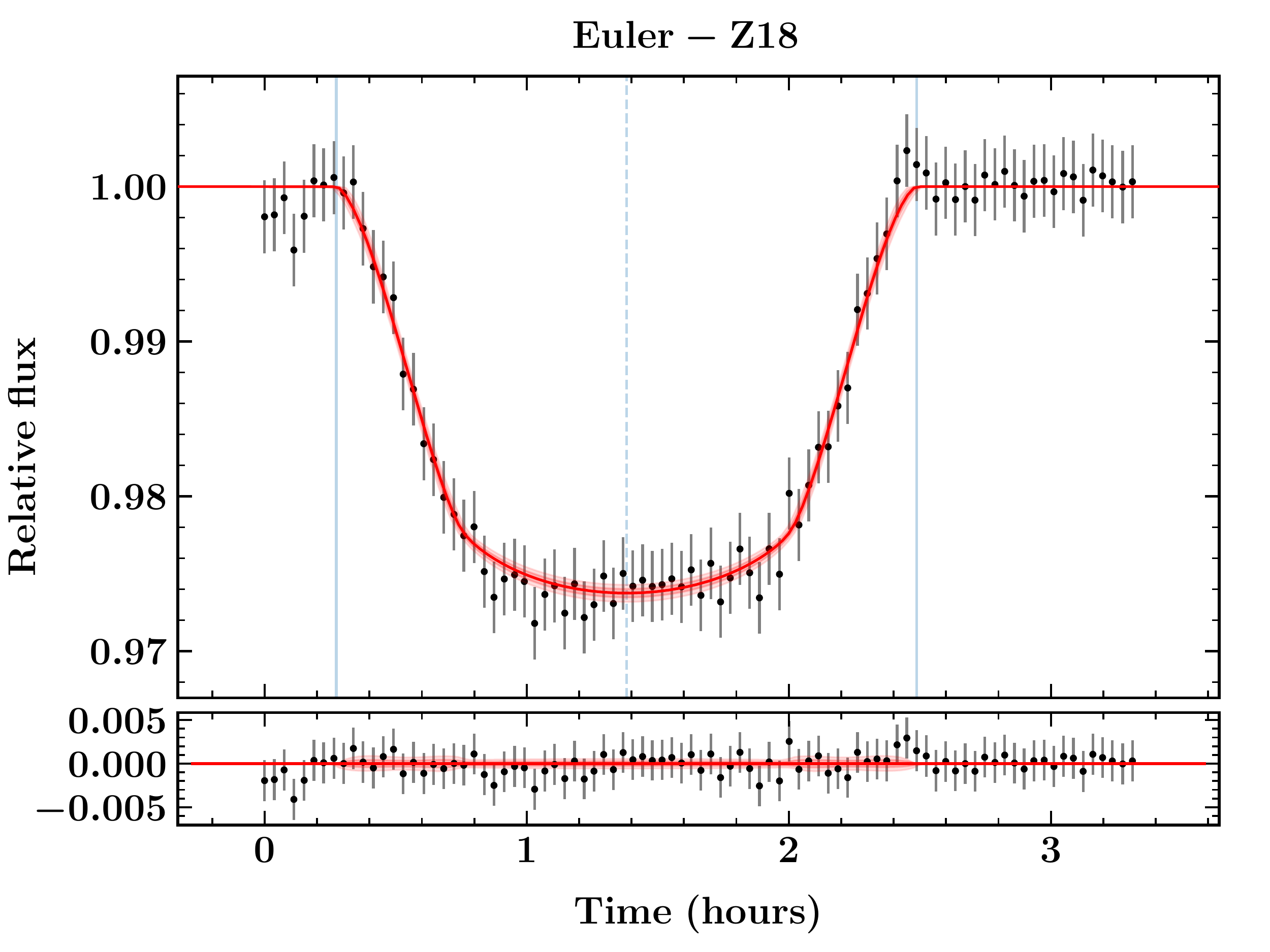}\\
        \rotatebox{00}{\tiny Time (hours)}
    \caption{Photometric follow-up light curves from SAAO \ref{saao} taken in the night 2017 March 31 (top left) and Euler \ref{euler}. The Euler light curves were taken in the night 2017 July 13 (top right), 2017 July 23 (bottom left), and 2018 March 8 (bottom right). The red curve and the pink shaded regions show the best model fit and the 1 and 2\,$\sigma$ confidence intervals (same as described for Fig.~\ref{fig::ngtsphase}).}
    \label{fig::photfu}
\end{figure}

To constrain the ephemeris of the planet, candidate photometric follow-up observations at the South African Astronomical Observatory (SAAO) and the La Silla Observatory (Chile) were  carried out. The observations are shown in Fig. \ref{fig::photfu}. 

\subsubsection{SAAO}\label{saao}
The transit of \planetname{} was observed through a Bessel V-band filter on the night of 2017 March 31 with the SHOC’n’awe camera, one of the frame-transfer CCD Sutherland High-speed Optical Cameras \citep{Coppejans2013} mounted on the 1m Elizabeth telescope at SAAO. The pixel scale of $0.167$\arcsec/pixel of these cameras is unnecessarily fine for our observations, therefore we binned  $4\times4$ pixels in the X and Y directions. All observations were made in focus. A total of 900 frames with exposure times of 20 seconds were obtained over the course of the night. Calibration frames for the data reduction were taken at sunset and sunrise. To achieve a high precision, we combined the calibration frames with
frames from the remaining observing run and performed bias correction and
flat-fielding of the data. We used a background correction for the science frames, and differential photometry was performed using a 5.1~pixel aperture and three fainter comparison stars that are visible in the $2.85'\times 2.85$\arcmin field of view. The comparison stars and aperture radius were chosen to maximise the signal-to-noise ratio of the light curve. All photometry was completed using the SEP python package \citep{Barbary1996}, which is based on the core algorithms of the Source Extractor \citep{Bertin1996}.  

\subsubsection{Euler}\label{euler}
We observed three transits of \starname{} with the Eulercam on the 1.2\,m Euler Telescope  at La Silla \citep{Lendl2012} on the nights beginning 2017 July 13, 2017 July 23, and 2018 March 08. The first observation was made using a Cousins I filter, a defocus of 0.15\,mm, and an exposure time of 60 seconds. However, the first four exposures used an exposure time of 50 seconds and the following eight exposures used 70 seconds because the observations were fine-tuned to the target and conditions; there was some variable thin cloud during the night. The latter two observations used a Gunn z-filter and no defocus. The exposure time on 2017 July 23 was 90 seconds (with two exposures of 100 seconds and two exposures of 120 seconds for fine-tuning), while on 2017 July 23, we used 120 seconds for the duration. The gap of $\sim$ 2 minutes in the light curve on 2017 July 23 is due to an error in the telescope control software. 

In all cases, the data were reduced using the standard procedure of bias subtraction and flat-field correction. Aperture photometry was performed with the PyRAF implementation of the phot routine. PyRAF was also used to extract information that was useful for de-trending; x- and y-position, FWHM, airmass, and sky background of the target star. The comparison stars and the photometric aperture radius were chosen in order to minimise the RMS scatter in the out-of-transit %.
portion of the light curve.

\subsection{Spectroscopic follow-up}
Initial vetting of \starname{} was performed with the CORALIE spectrograph \citep{Queloz2001} on the Swiss 1.2\,m telescope at La Silla Observatory, Chile. After checking for spectral binaries and blending scenarios, we obtained multi-epoch spectroscopy with the HARPS spectrograph \citep{2003Msngr.114...20M} on the ESO 3.6\,m telescope at La Silla Observatory, Chile, between March 2017 and May 2018 under programs \noprog. 

Radial velocities were calculated with a binary G2 mask using the standard data reduction pipelines for CORALIE and HARPS. The results are presented in  Table \ref{rvs}. Initial analysis confirmed the planetary nature of the candidate, showing radial velocity variations in phase with the NGTS ephemerides with a semi-amplitude of $K \sim 40 ~\ms$. A single measurement was corrected for moon-light contamination using the simultaneous sky-fibre. 

\begin{figure}
        \centering
        \includegraphics[width=\columnwidth]{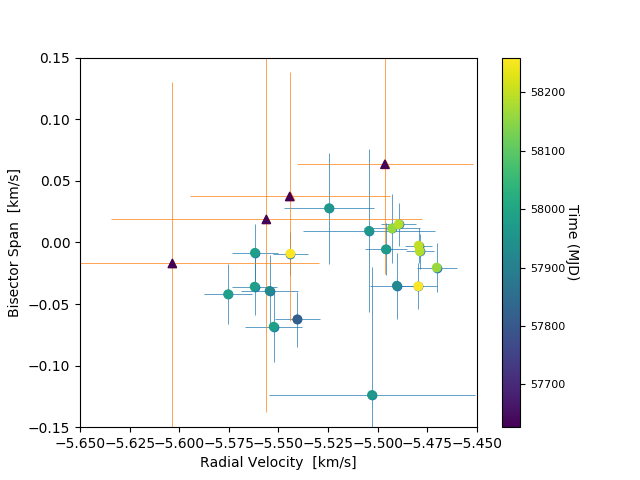}
    \caption{Bisector span over radial velocity measurements with HARPS (blue error bars) and CORALIE (orange error bars). No correlation is visible.}
    \label{fig::bis}
\end{figure}

To ensure that the radial velocity signal does not originate from cool stellar spots or a blended eclipsing binary, we checked for correlations between the line bisector spans \citep{QuelozBIS} and the radial velocity measurements. We find no evidence for a correlation; see Fig.~\ref{fig::bis}.
In order to characterise the stellar properties of \starname,{} we combine and shifted the 17 HARPS spectra in wavelength that were not contaminated by moon light to create a spectrum with a high signal-to-noise ratio for the analysis in Section~\ref{sub:stellar}.

\begin{table}
\caption{Coralie and HARPS radial velocity measurements of \starname.\label{rvs}}
\begin{tabular}{lccr}
\hline
\hline
BJD$_\mathrm{TDB}$ & Radial Velocity & $\sigma_{\mathrm{RV}}$ &   Instr. \\
$-$2,450,000 & (\kms) & (\kms)  &  \\
\hline
\noalign{\smallskip}
7626.507247     & -5.593    & 0.051     &  Coralie\\
7630.508626     & -5.652    & 0.074     &  Coralie\\
7631.486303     & -5.545    & 0.044     &  Coralie\\
7632.478367     & -5.605    & 0.078     &  Coralie\\
7814.823955     &  -5.5404      &  0.0113 &  HARPS\\
7924.610489     &  -5.4903      &  0.0134 &  HARPS\\
7925.621469     &  -5.5545      &  0.0145 &  HARPS\\
7960.533932     &  -5.5245      &  0.0225 &  HARPS\\
7961.544233$^{*}$       &  -5.5028      &  0.0519 &  HARPS\\
7964.537489     &  -5.5043      &  0.0331 &  HARPS\\
7974.501742     &  -5.4958      &  0.0106 &  HARPS\\
7979.480682     &  -5.5620      &  0.0114 &  HARPS\\
7980.488371     &  -5.5524      &  0.0145 &  HARPS\\
7982.484781     &  -5.5618      &  0.0117 &  HARPS\\
7983.480932     &  -5.5754      &  0.0121 &  HARPS\\
8159.833208     &  -5.4928      &  0.0140 &  HARPS\\
8162.826420     &  -5.4702      &  0.0101 &  HARPS\\
8188.813394     &  -5.4893      &  0.0087 &  HARPS\\
8189.814538     &  -5.4788      &  0.0071 &  HARPS\\
8199.758555     &  -5.4793      &  0.0069 &  HARPS\\
8258.758482     &  -5.5439      &  0.0088 &  HARPS\\
8259.665546     &  -5.4795      &  0.0094 &  HARPS\\
\hline
\end{tabular}
\begin{list}{}{\leftmargin=0em \itemindent=1.20em}
 \item[* Corrected for moon-light contamination]
 \end{list}
\end{table}

\section{Analysis}
\label{sec::ana}
\subsection{Stellar classification}\label{sub:stellar}
\begin{figure}
        \centering
        \includegraphics[width=0.95\columnwidth]{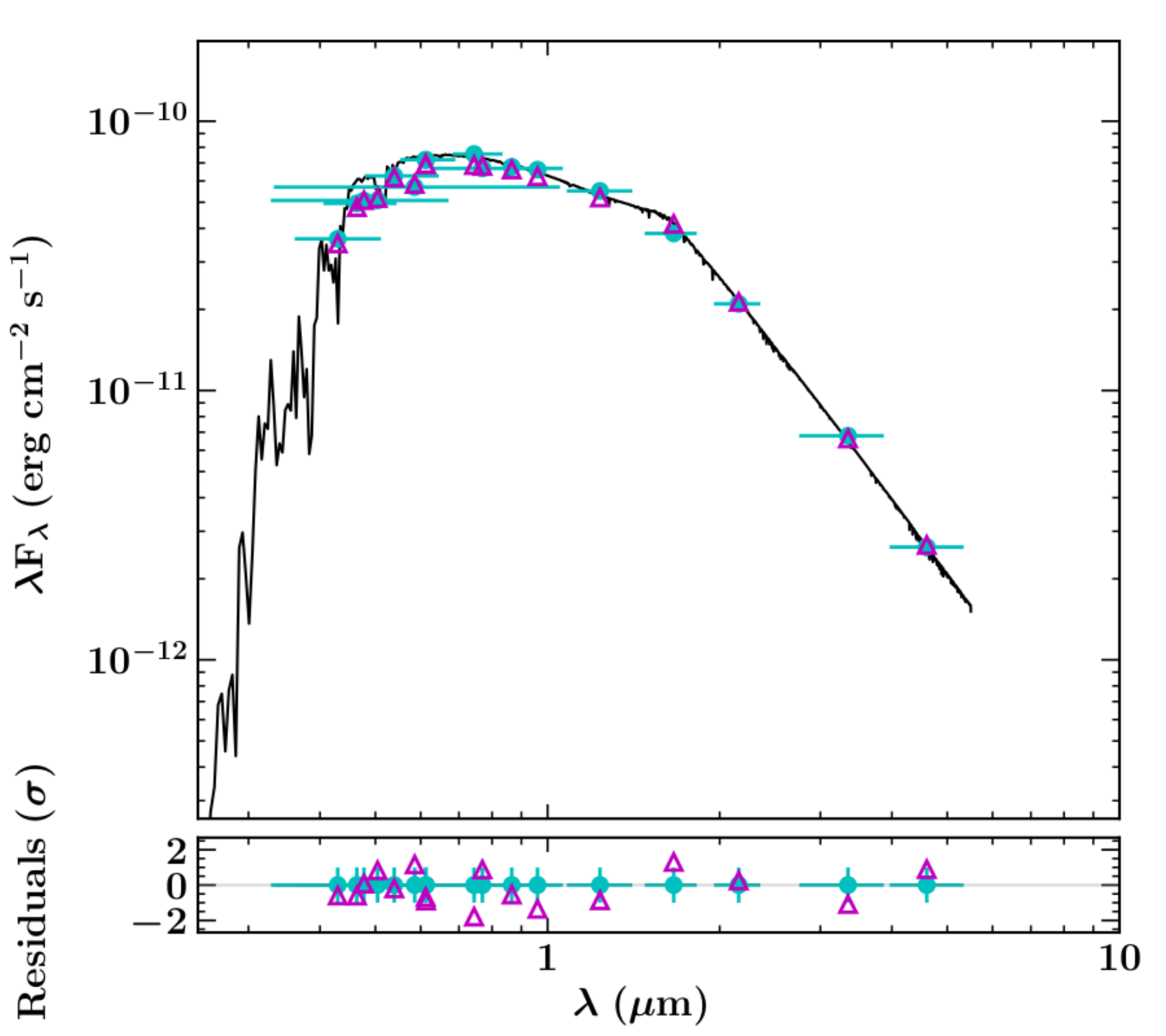}
    \caption{Best-fitting PHOENIX model SED (black line) for \starname{} based on the photometric data (cyan points) presented in Table \ref{tab:stellar}. The horizontal cyan lines indicate the widths of the different photometric bands. The magenta triangles represent the model flux convolved with each bandpass. Bottom: Residuals of the fit in units of observational uncertainty.}
    \label{fig:sed}
\end{figure}

With data from the GAIA DR2, %GAIA DR2 data release 
the stellar mass and radius can now be determined directly from the
bolometric flux, effective temperature, and parallax. Applying the method described in \citet{Gillen2017}, we used broadband photometric magnitudes from different catalogues as given in Table \ref{tab:stellar} \citep{APASS,SDSS,Twomass,GAIA2018,Wise,PanStarrs} together with astrometric information from GAIA DR2 \citep{GAIA2018} to determine the stellar parameters to high accuracy. The spectral energy distribution (SED) was modelled by convolving PHOENIX \citep{Phoenix} and BT-SETTL \citep{BT-Settl} models using Markov chain Monte Carlo (MCMC) simulations. 
The best-fitting PHOENIX model is shown in Fig. \ref{fig:sed}.

The resulting stellar parameters are in agreement with those that were determined using spectral classifications on the calibrated HARPS spectra, taken for radial velocity follow-up. For comparison we used a standard spectral classification following a method similar to the method reported by \citet{Doyle2013} and the Empirical SpecMatch tool as described in \citet{2017ApJ...836...77Y}. The results from all three methods agree with each other.\\

%To get the most accurate stellar parameters the very precise radius estimate from the above described method is used in combination with the stellar density as is it retrieved from the light curve analysis described in section \ref{sec:mod} we determine the \logg and stellar mass.
%AC: was a little confused by the above sentence, so double check that I got it right!
To obtain the most accurate stellar parameters, we determined the %\logg and 
stellar mass by combining the very precise radius estimate from the above method with the stellar density as retrieved from the light curve analysis described in section \ref{sec:mod}. The host star is a K2 dwarf with a radius of \Rstar\,=\,\PrimaryRadius~\Rsun, a mass of \Mstar\,=\,\PrimaryMass~\Msun \, , and an effective temperature of \teff\,=\,\PrimaryTeff~K. The catalogue information for the host star and the stellar parameters is listed in Table \ref{tab:stellar}. We analysed the stellar spectra and photometric data for signs of stellar activity. The signal-to-noise ratio of the spectrum is too  around the Ca H+K lines to conclude about stellar activity. There are no emission lines above the noise. The H-alpha profiles look normal, with no obvious signs of any emission or core infilling. The photometric data show no indications of stellar activity either.

\begin{table*}[!th]
\begin{center}
\caption{Main identifiers, coordinates, magnitudes, and spectroscopic parameters of \starname.\label{tab:stellar}}
\begin{tabular}{lcc}
\hline
\hline
\noalign{\smallskip}
Parameter & {\texttt{\starname}} & Unit\\
\noalign{\smallskip}
\hline
\noalign{\smallskip}
RAJ2000 (GAIA DR1)      & 14$^h$44$^m$13$^s$.9842   & h \\
DECJ2000 (GAIA DR1) & +05$\degr$36$^\prime$19$^{\prime \prime}$.347  & deg\\
%GAIA DR1 source ID & 1159042829436840832 & \ldots\\
GAIA DR2 source ID & 1159042833731813760 & \ldots\\
Parallax & $3.2310 \pm 0.0272$ & $\mu as$\\
2MASS ID &  14441396+0536195     & \ldots\\
\noalign{\smallskip}
\hline
\noalign{\smallskip}
Effective temperature \teff & \PrimaryTeff  & K\\
Surface gravity \logg & \PrimaryLogglc  & cgs\\
Metallicity [Fe/H] & \PrimaryMet    & dex \\
%\vmic & $0.95 \pm 0.03$   & \kms &\\
%\vsini & \Primaryvsini &  \kms \\
%Age                            & \PrimaryAge                 &Gyr\\
Spectral type &  K2\,V & \ldots\\
\hline
\noalign{\smallskip}
Primary mass $M_\ast$                           & \PrimaryMass           &\Msun\\
Primary radius $R_\ast$                         & \PrimaryRadius            &\Rsun\\
\noalign{\smallskip}
\hline
\noalign{\smallskip}
APASS B & $14.706 \pm 0.052$ & mag\\
         APASS V  & $13.770  \pm  0.054$ & mag\\
         SDSS g   & $14.189  \pm  0.032$ & mag\\
         SDSS r   & $13.483  \pm  0.031$ & mag\\
         SDSS i   & $13.218   \pm 0.024$ & mag\\
         PSg\_AB   & $14.132   \pm 0.003$ & mag\\
         PSr\_AB   & $13.484  \pm  0.000$ & mag\\
         PSz\_AB   & $13.173  \pm  0.010$ & mag\\
         PSy\_AB   & $13.086  \pm  0.004$ & mag\\
         Gaia G   & $13.526  \pm  0.002$ & mag\\
         Gaia BP  & $14.024  \pm  0.002$ & mag\\
         Gaia RP  & $12.893  \pm  0.003$ & mag\\
         2MASS J  & $12.117  \pm  0.023$ & mag\\
         2MASS H  & $11.708  \pm  0.027$ & mag\\
         2MASS Ks & $11.612  \pm  0.025$ & mag\\
         WISE-1   & $11.526  \pm  0.023$ & mag\\
         WISE-2   & $11.575  \pm  0.021$ & mag\\
     \noalign{\smallskip}
\hline
\end{tabular}
\end{center}
\end{table*}

\subsection{Combined modelling}\label{sec:mod}
We modelled the light curves from NGTS and photometric follow-up observations together with the radial velocity data obtained at HARPS using \gpe\ \citep{Gillen2017}. 
The radial velocity data showed a long-term trend in addition to the expected Keplerian orbit modulation. Furthermore, the CORALIE and HARPS data did not overlap in the timing of their observations; the CORALIE data were taken $\text{about }$six months before the HARPS observations. Because of the long-term trend, an unknown radial velocity offset between the zero-points of the two spectrographs, and large uncertainties on the CORALIE data relative to the semi-amplitude of the orbital modulation, we opted to remove the CORALIE data from our global modelling. Including them would have added model parameters and a correlation between the spectrograph zero-point offsets and the long-term trend with no obvious gain in radial velocity constraint.
The main parameters we modelled were the sum of radii relative to the semi-major axis of system, the radius ratio, the cosine of the orbital inclination, the orbital period and epoch, the systemic radial velocity, the semi-amplitude of the radial velocity, and the sine of the linear radial velocity trend. 
For the limb darkening we assumed a quadratic law (using LDtk; \citealt{Parviainen15}), which gave two additional free parameters for each photometric instrument (NGTS, SAAO, and EulerCam).
The light curves with the best-fit model are shown in Figs. \ref{fig::ngtsphase} and \ref{fig::photfu}. We observe no transit timing variation of the transits.\\

We also modelled the system while allowing the eccentricity to vary.
We found a best-fit eccentricity of $0.18 \pm 0.11$. An additional test, fitting the data allowing for eccentricity but not for a linear trend gave us an eccentricity of $0.11 \pm 0.08$. Because the observations are limited in number, we found that the best-fit eccentricity was not significant and therefore decided for the final model to fix the system on a circular orbit.  The implications for the radius and mass of the planetary companion were negligible in either case.

\begin{figure}
        \centering
        \includegraphics[width=0.95\columnwidth]{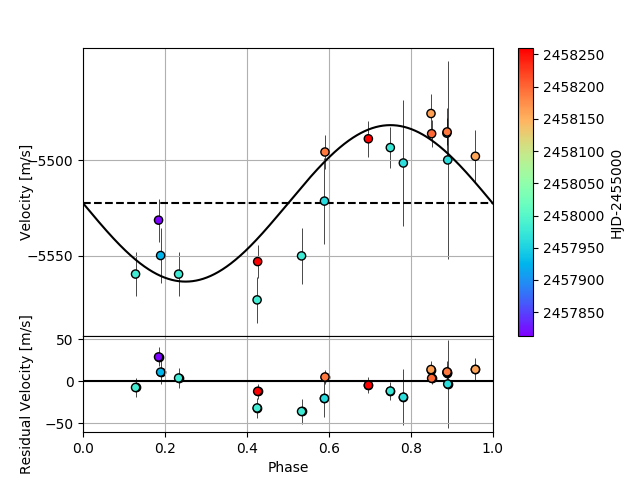}
    \caption{Modelled radial velocity measurements in orbital phase, colour-coded by the time of observation. The fit of the best orbital solution is given by the black line.}
    \label{fig::rv}
\end{figure}

\begin{figure*}
        \centering
        \includegraphics[width=1.\textwidth]{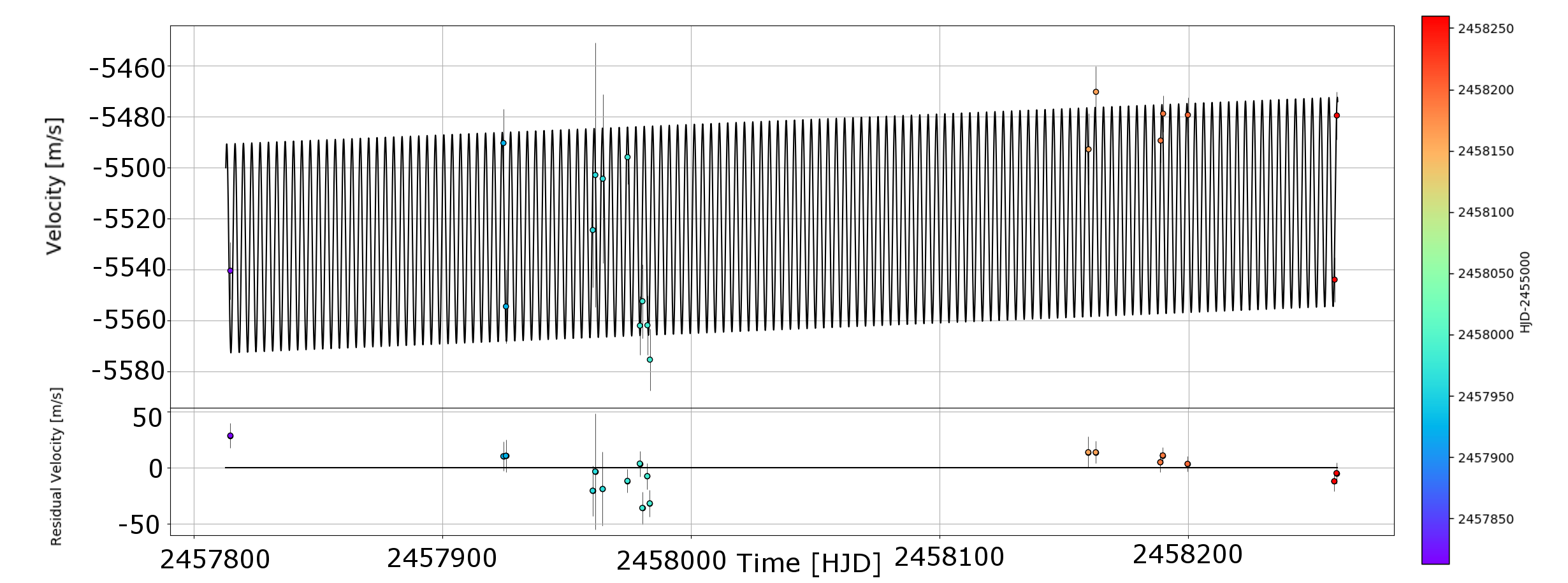}
    \caption{Radial velocity measurements over time, colour-coded by the time of observation. The fit of the best orbital solution including a linear trend} is given by the black line.
    \label{fig::rv2}
\end{figure*}

The residuals in the fit to the radial velocity data suggest that the assumed linear trend does not describe the observational data well. Fitting a second Keplerian solution to the residuals would be over-fitting the available data, however.
The best-fit model together with the radial velocity data is shown in Figure \ref{fig::rv} (phase folded) and Figure \ref{fig::rv2} (over time).

\begin{figure}
        \centering
        \includegraphics[width=\columnwidth]{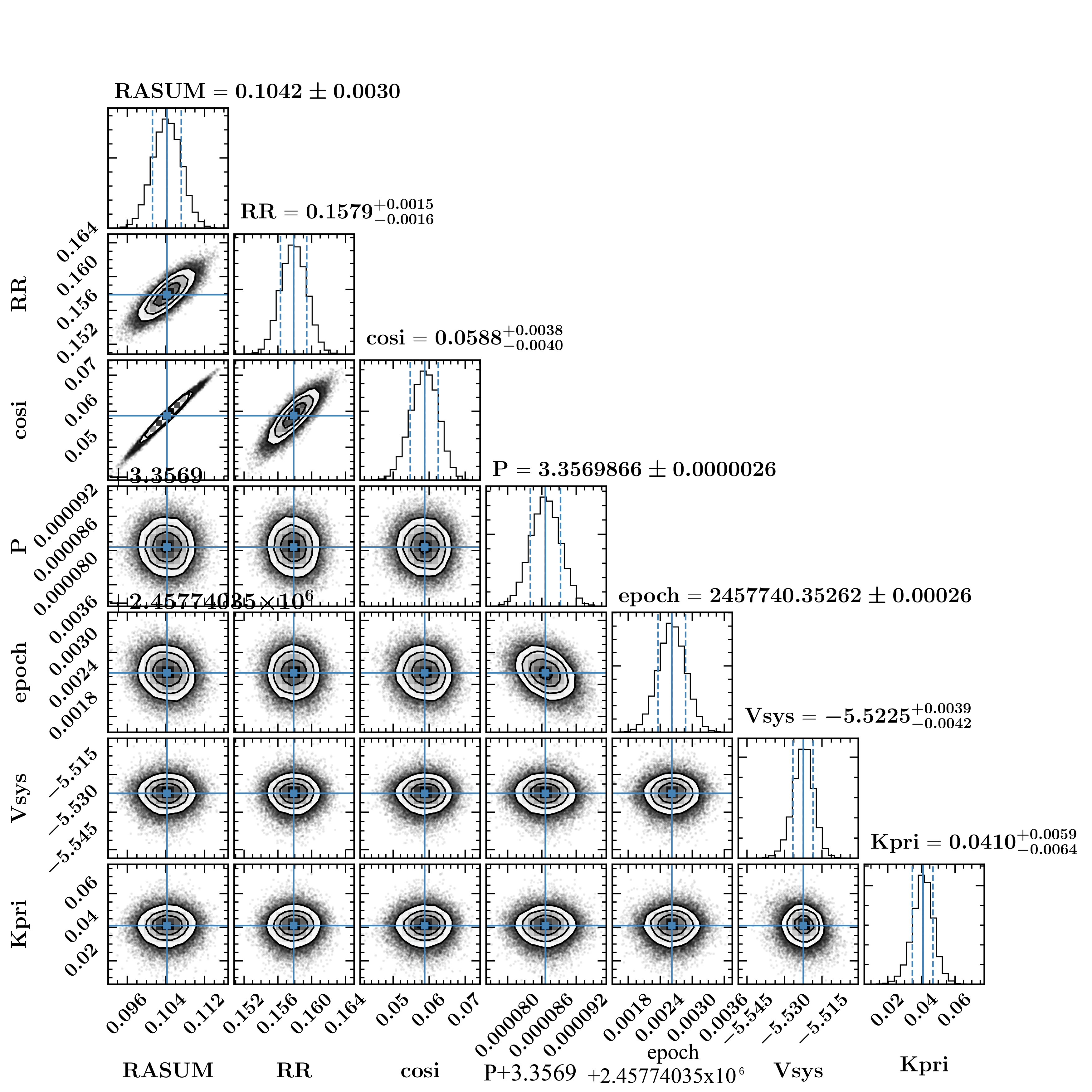}
    \caption{MCMC posterior distribution for the main free parameters.}
    \label{fig::distr}
\end{figure}

The MCMC posterior distribution for the main parameters is shown in Figure \ref{fig::distr}.

\section{Results}\label{sec::res}
The K2 main-sequence star NGTS-5 is orbited by short-period exoplanet. \planetname{} transits its star every \PERIOD \, days. The orbit is slightly inclined with an inclination of i=\INCLINATION \, degrees.
The eccentricity is not well constrained from our radial velocity follow-up observations. We therefore assumed the orbit to be circular. The residuals in the radial velocity data might be due to an additional orbiting element in the system, but the data at hand do not allow us to restrict its parameters.  For our final modelling we applied a linear trend to account for the additional signal. The residuals in Figure \ref{fig::rv2} show that a linear trend might not be the best explanation for the additional signal, but more free parameters would lead to over-fitting the data.  To understand the implications to the bulk properties of the planet, we modelled the system with eccentricity as a free parameter, as well as without any additional trend and assuming a quadratic trend. The bulk parameters of the planet did not change significantly and were consistent within 1 $\sigma$.\\ 

The transit light curves allow us to constrain that the radius of the planet is slightly larger than the radius of Jupiter with $R_P$=\SecondaryRadius \,\Rjup. The radial velocity follow-up measurements, however, show the mass to be significantly lower than the mass of Jupiter, $M_P$=\SecondaryMass \,\Mjup, resulting in an extremely low mean density of $0.19 g/cm^3$.

\begin{table*}[!th]
\begin{center}
\caption{Parameters from light curve and radial velocity data analysis.\label{tab:param}}
\begin{tabular}{lcc}
\hline
\hline
\noalign{\smallskip}
Parameter & combined modelling & Unit\\
\noalign{\smallskip}
\hline
\noalign{\smallskip}
Orbital period $P_\mathrm{orb}$  &              \PERIOD  & days\\
Transit epoch $T_0$                      &              \EPOCH   &BJD$_\mathrm{TDB}-2450000$\\
Transit duration                                 &              \DURATION                        & hours\\
Scaled semi-major axis $a/R_*$   &      \SSMA   &\\
Semi-major axis $a$                      &    \SMA        & au\\
Scaled secondary radius $R_2/R_1$        &  \SSR  &\\
Orbital inclination angle $i$    &    \INCLINATION    &$\deg$\\
%Impact parameter $b$                    &    \IMPACT           & \\
Limb-darkening coefficient z$u_+$ &   \LDOnez &\\
Limb-darkening coefficient z$u_-$ &   \LDTwoz  &\\
Limb-darkening coefficient I$u_+$ &   \LDOnei &\\
Limb-darkening coefficient I$u_-$ &   \LDTwoi  &\\
Limb-darkening coefficient V$u_+$ &   \LDOnev &\\
Limb-darkening coefficient V$u_-$ &   \LDTwov  &\\
Limb-darkening coefficient NGTS$u_+$ &   \LDOnen &\\
Limb-darkening coefficient NGTS$u_-$ &   \LDTwon  &\\
%Surface Gravity \logg\footnote{Stellar surface gravity as resulting from light curve modeling.} & \PrimaryLogglc & dex \\ 
\noalign{\smallskip}
\hline
\noalign{\smallskip}
Radial velocity semi-amplitude $K$ &  \KAmp  &$ \ms $\\
Systemic radial velocity $\gamma$ & \GAMMA &  \kms  \\
Sine of the linear radial velocity trend & $0.015 \pm 0.012$ &  [km\,$s^{-1}$\,$yr^{-1}$]\\
Eccentricity $e$ & \ECCENTRICITY                       &\\
\noalign{\smallskip}
%$M^{1/3}_1/R_1$                                        &       \MoverR         & &Solar units\\
%Primary mean density $\rho_1$ & \PrimaryRho                             & &\gcm \\
%Primary surface gravity \logg\footnote{\label{note}Derived from the light curve modeling, effective temperature, metal content, and isochrones.} & \PrimaryLogg   & &cgs\\
%Distance               & & & parsec\\
\noalign{\smallskip}
\hline
\noalign{\smallskip}
Secondary mass $M_P$                    & \SecondaryMass   &$M_\mathrm{Jup}$\\
Secondary radius $R_P$                  & \SecondaryRadius  &$R_\mathrm{Jup}$ \\
%Secondary mean density          & & &\gcm\\
Secondary surface gravity \loggp                                & \SecondaryLogg   &cgs\\
Secondary calculated effective temperature & \SecondaryTeff  &K\\
\noalign{\smallskip}
\hline
\end{tabular}
\end{center}
\end{table*}

\section{Discussion}\label{sec::dis}
\begin{figure*}
        \begin{subfigure}[t]{\columnwidth}
        \includegraphics[width=0.95\columnwidth]{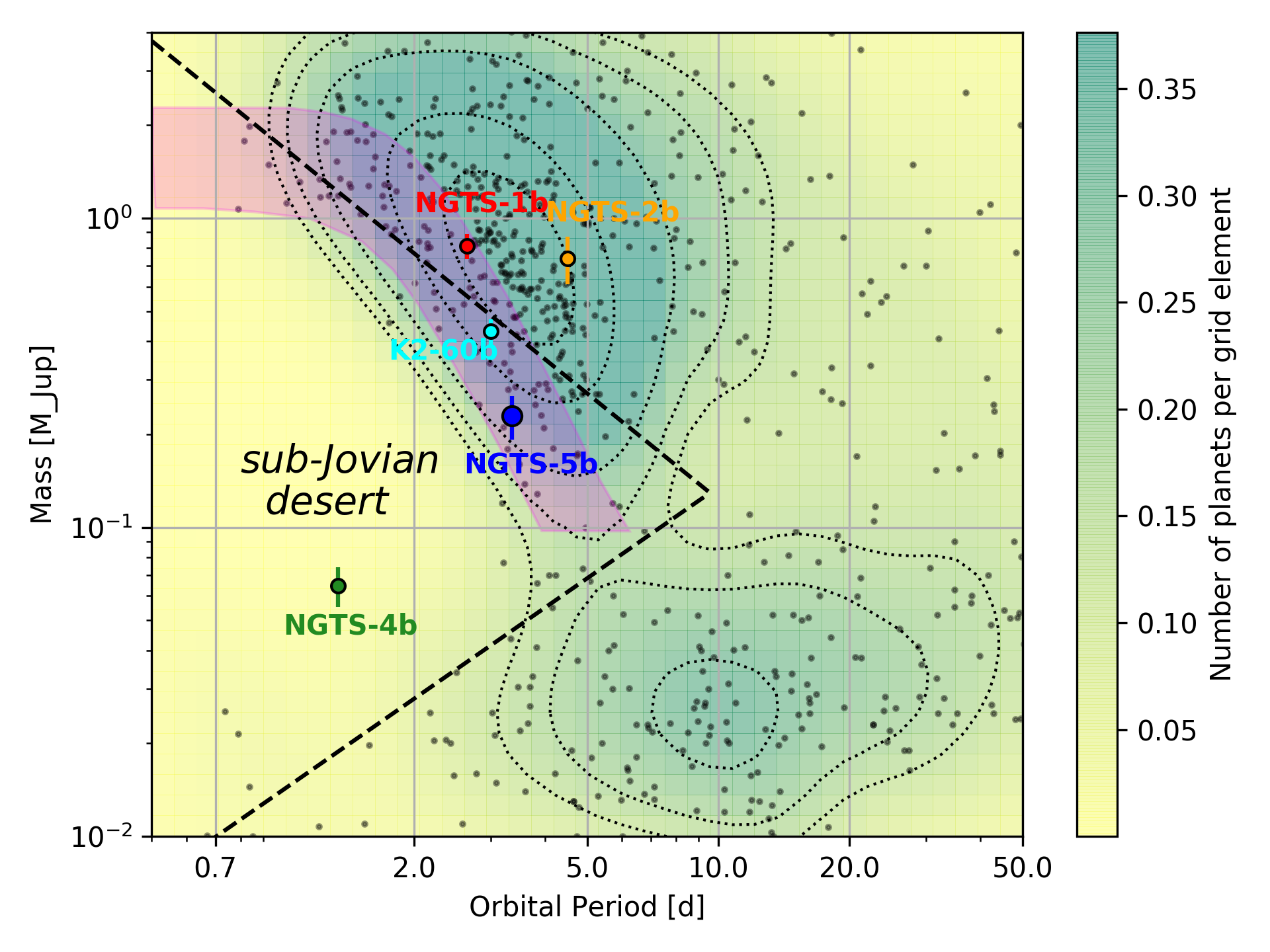}
    \newsubcap{Planetary mass over orbital period. Plotted are all stars from the NASA Exoplanet Archive with a radius uncertainty lower than 10\% or a mass uncertainty lower than 50\%. The blue point shows \planetname. Additionally, we highlight NGTS-1b in red \citep{Bayliss2018},  NGTS-2b in orange \citep{Raynard2018}, NGTS-4b in green \citep{West2018}, and K2-60b  in cyan \citep{Eigmuller2017}. The black dashed lines show the boundaries of the sub-Jovian desert as determined in \citet{Mazeh2016}. The magenta region shows the high-eccentricity  migration boundary including tidal decay as presented in \citet{Owen2018}. The background and the dotted black lines highlight the point density per grid element (the plot is divided into 40 x 40 equally spaced grid elements) of our sample. The yellow region on the left shows the sub-Jovian desert.}
    \label{fig::subjov_mazeh1}
    \end{subfigure}\hfill
    \begin{subfigure}[t]{\columnwidth}
        \includegraphics[width=0.95\columnwidth]{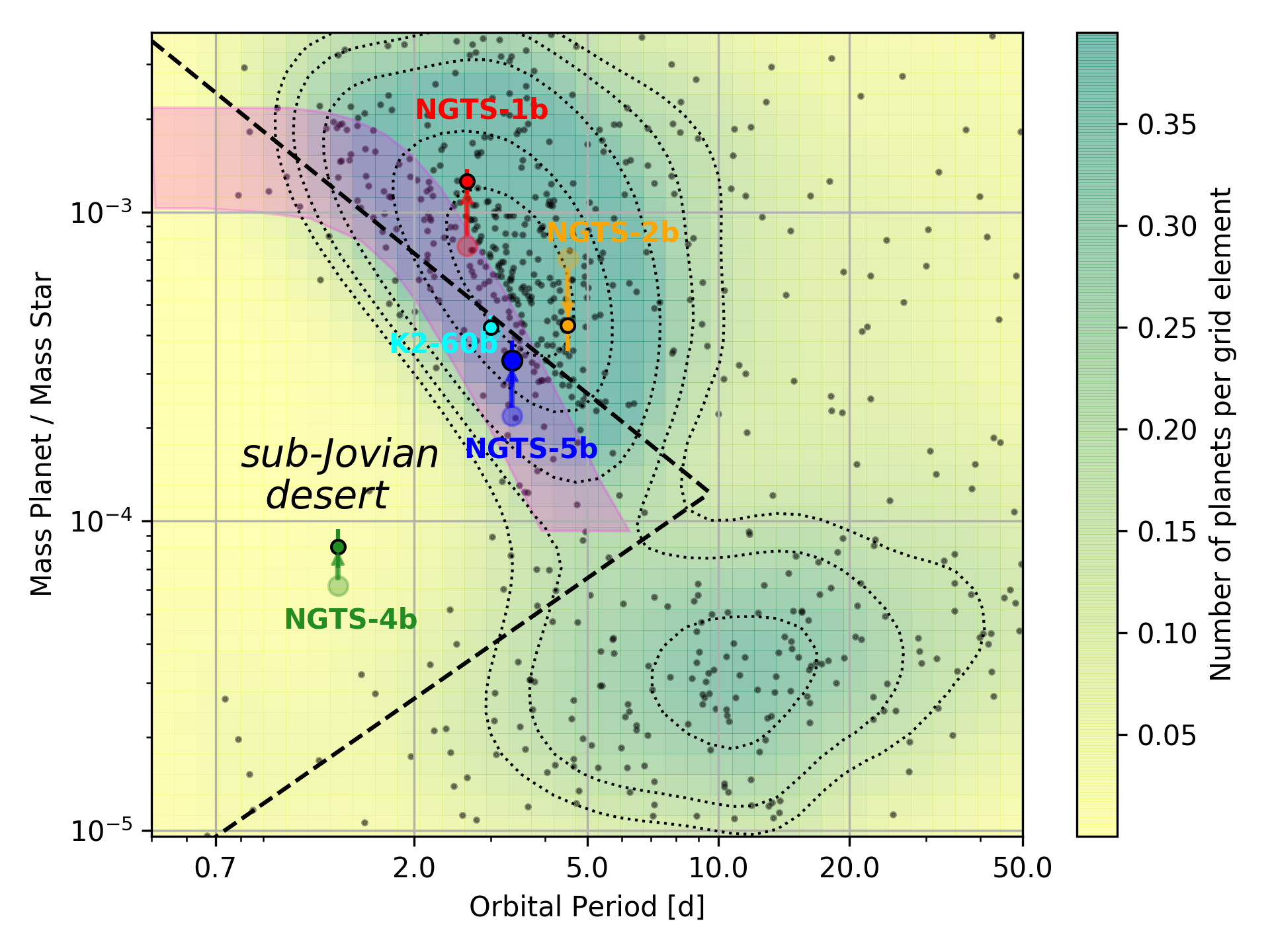}
    \newsubcap{Mass ratio (planetary mass over stellar mass) plotted over the orbital period.  The data are the same as in Fig.\ref{fig::subjov_mazeh1}. The black dashed lines and magenta region show the boundaries of the sub-Jovian desert as in Fig. \ref{fig::subjov_mazeh1}, but converted into the mass ratio assuming solar mass for the host star. Highlighted are \planetname{} (blue, K-dwarf host), NGTS-1b (red, M-dwarf host), NGTS-2b (orange, F-dwarf host), NGTS-4b (green, K-dwarf host), and K2-60b (cyan, G-dwarf host). To highlight the influence of the stellar type, we also show for these stars the position in the diagram; the sun is also assumed to be a host star (i.e. stellar parameters are not taken into account; shaded). The arrows show the change that is effected  when the host star is taken into account.}
    \label{fig::subjov1}
    \end{subfigure}
\end{figure*}

\begin{figure*}
        \begin{subfigure}[t]{\columnwidth}
        \includegraphics[width=0.95\columnwidth]{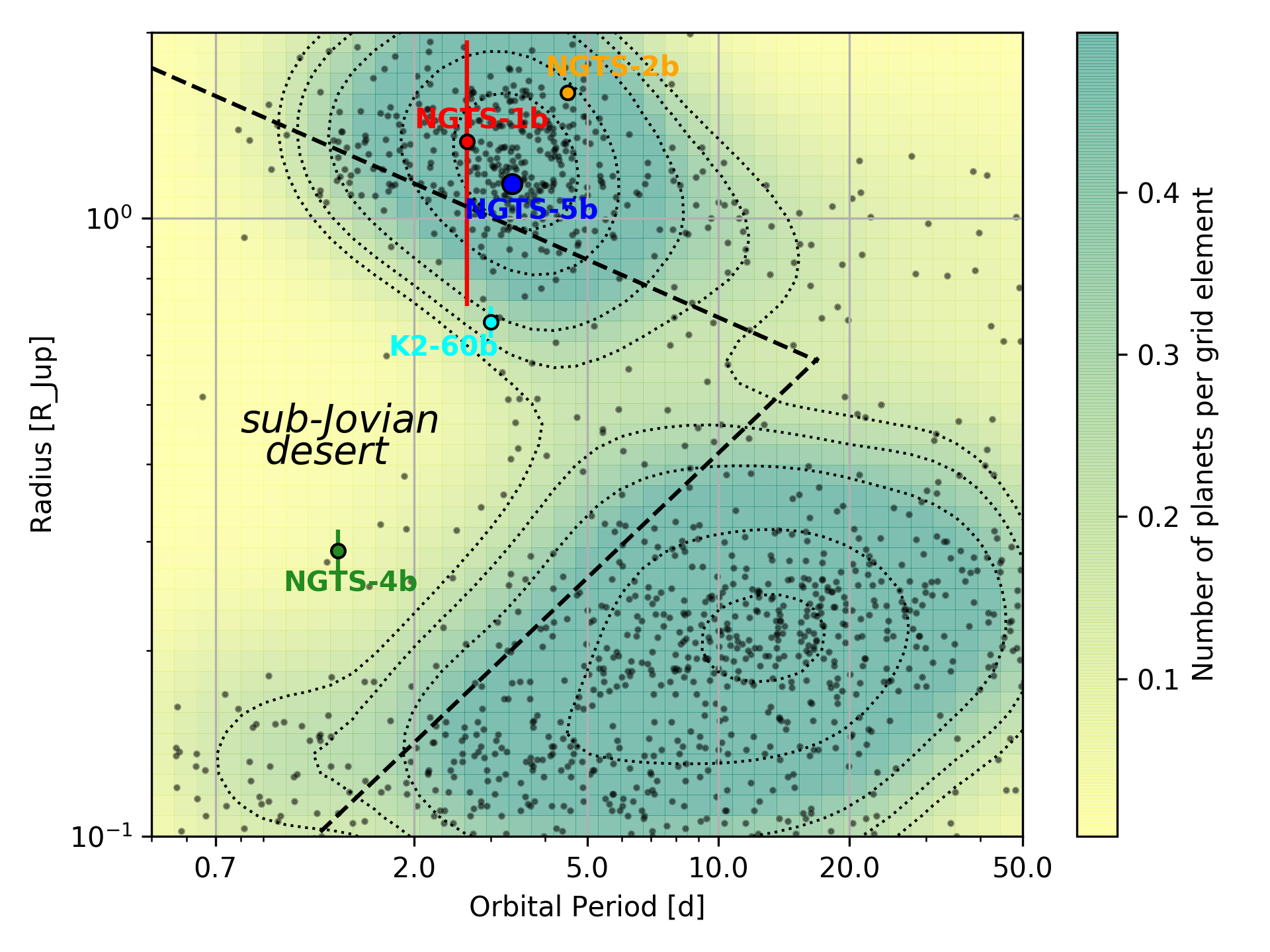}
    \newsubcap{Same as Fig.\ref{fig::subjov_mazeh1}, but for the planetary radius over orbital period.}
    \label{fig::subjov_mazeh2}
    \end{subfigure}\hfill
    \begin{subfigure}[t]{\columnwidth}
    \includegraphics[width=0.95\columnwidth]{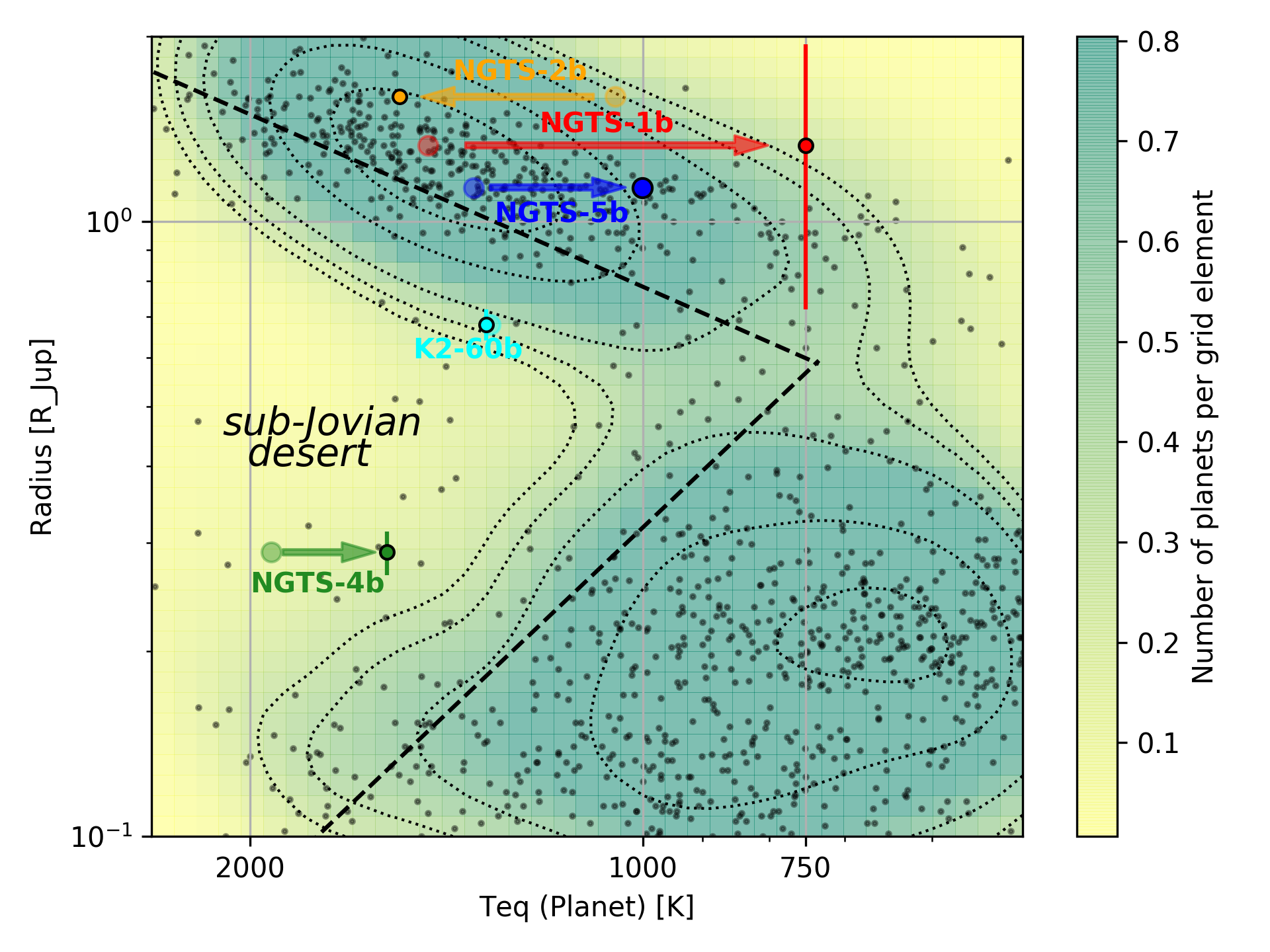}
    \newsubcap{Same as Fig. \ref{fig::subjov1}, but for the planetary radius over equilibrium temperature.}
    \label{fig::subjov2}
    \end{subfigure}
\end{figure*}

With its short orbital period and low mass, \planetname{} lies within the mass/period sub-Jovian desert as defined in \citet{Mazeh2016}. In Figure \ref{fig::subjov_mazeh1}, \planetname{} is plotted (blue) together with other planets selected from the NASA Exoplanet Archive\footnote{The data we used include all confirmed planets with an uncertainty in radius lower than 10\% and an uncertainty in mass lower than 50\% as available on 18 April 2019.} \citep{nasaexoplanet} (coloured points highlight planets which are used later when discussing Figure \ref{fig::subjov1} and Figure \ref{fig::subjov2}). The black dashed lines correspond to the empirically defined boundaries of the sub-Jovian desert \citep{Mazeh2016}. The background and the dotted line represent the point density present in our planet sample. The magenta region in Fig. \ref{fig::subjov_mazeh1} represents the high-eccentricity  migration  boundary including tidal decay as presented in \citet{Owen2018}. In comparison with the boundary from \citet{Mazeh2016} \planetname{} lays within the desert, whereas the theoretical boundary as well as the point density of current exoplanet population suggest \planetname{}  to be closer to the upper boundary of the sub-Jovian desert.  

In the radius-period plane (Fig. \ref{fig::subjov_mazeh2}) \planetname{} is well above the boundary presented by \citet{Mazeh2016} and lies in the middle of the population of inflated hot Jupiters.

Figures \ref{fig::subjov_mazeh1} and \ref{fig::subjov_mazeh2} display the sub-Jovian desert as it is typically shown \citep[e.g.][]{Mazeh2016,Matsakos2016,Owen2018}. This however does not take the stellar information of the planet host into account, but instead assumes all host stars to be solar like.  
As \starname{} cannot be approximated with solar parameters, our conception of \planetname{} in regards to the sub-Jovian desert, as depicted in  Figures \ref{fig::subjov_mazeh1} and \ref{fig::subjov_mazeh2} may be misleading. A recent study by \citet{szabo2019}, which was published during our reviewing process, also shows the dependence of the sub-Jovian desert on stellar parameters. Processes thought to be responsible for shaping the sub-Jovian desert are related to the host stars' parameters. One effect which may be responsible for shaping the upper boundary of the sub-Jovian desert is inflation caused by e.g. ohmic heating. This effect is not directly correlated to the orbital period. The direct dependence of inflation would be to the insolation or equilibrium temperature of the planet. By using the planet's equilibrium temperature instead of the orbital period, we thus show the dependency caused by inflation when the host star is taken into account. Figure \ref{fig::subjov2} can be directly compared to Fig. \ref{fig::subjov_mazeh2}. The y-axis is the same, but for the x-axis, the orbital period has been replaced by the planetary equilibrium temperature assuming an albedo of 0.2 for all planets. For \planetname{}, K2-60b, NGTS-1b, NGTS-2b, and NGTS-4b, the difference between assuming solar properties for the host star (shaded) and using the actual host star parameters is highlighted with arrows. The properties of these planets are listed in Table \ref{tab::planets}. This difference directly shows the importance of taking stellar properties into account depending on the stellar type. K2-60 is a G-type star, therefore changes are not significant. \planetname, NGTS-1b, and NGTS-4b all orbit K- or M-dwarfs and show a clear shift to lower equilibrium temperatures. NGTS-2, on the other hand, is an F-type star, which is the reason that we can see the opposite effect.  
In addition to the significant influence on single planets, such as \planetname, the shape of the hot-Jupiter population is also affected in general. A comparison of Figures \ref{fig::subjov_mazeh2} and \ref{fig::subjov2} shows that the equilibrium temperature is more strongly dependent on the radius at the upper boundary of the sub-Jovian desert. The slope of the contour lines below the population of the hot Jupiters, which is indicative of the upper boundary of the sub-Jovian desert, also changes and agrees better with the results of \citet{Mazeh2016} when stellar type is accounted for (cf. Fig. \ref{fig::subjov2}). This shows how much more diverse our sample is than the sample that was analysed in \citet{Mazeh2016}.\\

Next to the inflation, which is related to the stellar insolation, another effect that shapes the desert is high-eccentricity migration. As a dynamical process, it is not directly dependent on the stellar insolation, but is instead directly correlated with the orbital period of the planet. However, the host star also plays a crucial role in this effect because the dynamical processes are correlated to the stellar mass. \citet{Matsakos2016} modelled only planets orbiting solar mass stars, which required them to take the host star into account when they displayed the sub-Jovian desert. Equations 3 and 4 in \citet{Matsakos2016} clearly show, however, that the mass of the host star should be taken into account in a more diverse sample. In Fig. \ref{fig::subjov1} we show the sub-Jovian desert by plotting the mass ratio over the orbital period. As in Fig. \ref{fig::subjov2}, we highlight the planets \planetname, K2-60b, NGTS-1b, NGTS-2b, and NGTS-4b and the influence of their host star on their position in this diagram in relation to the overall planet population.

\begin{savenotes}
\begin{table}
\begin{threeparttable}
\begin{center}\small
  \caption{Parameters (planetary radius, planetary mass, orbital period, equilibrium temperature, stellar radius, stellar mass, and stellar effective temperature for the selected planets used in Figures \ref{fig::subjov_mazeh1}, \ref{fig::subjov1}, \ref{fig::subjov_mazeh2}, and \ref{fig::subjov2}.}\label{tab::planets}

\begin{tabular}{c|c c c c c c c} 
    \hline
    \hline
    \noalign{\smallskip}
    Planet & $R_P$  & $M_P$  & P & $T_{eq}$& $R_*$  & $M_*$  & $T_{eff}$\\
     & [\Rjup{}] & [\Mjup{}] & [d] & [K] & [\Rsun{}] & [\Msun{}] & [K]\\
 \hline
 NGTS-1b & 1.33\tnote{*} & 0.81 & 2.65 & 750 & 0.57 & 0.62 & 3916\\ 
 NGTS-2b & 1.60 & 0.74 & 4.51 & 1535 & 1.70 & 1.64 & 6478\\
 NGTS-4b & 0.29 & 0.06 & 1.33 & 1572 & 0.84 & 0.75 & 5143\\
 NGTS-5b & 1.14 & 0.23 & 3.36 & 1001 & 0.74 & 0.66 & 4987\\
 K2-60b  & 0.68 & 0.43 & 3.00 & 1318 & 1.12 & 0.97 & 5500\\
\end{tabular}
\begin{tablenotes}\footnotesize
\item[*] NGTS-1b only shows grazing transits. The radius of the planet therefore has large uncertainties.
\end{tablenotes}
\end{center}
\end{threeparttable}

\end{table}
\end{savenotes}

%Taking the host star into account \planetname moves out of the sub-Jovian desert and is now at the upper boundary of the sub-Jovian desert and clearly not  within the desert. 
When the host star is taken into account, \planetname{} is shifted from within the desert to its upper boundary. NGTS-4b moves more into the middle of the desert, but NGTS-4b with its low mass might belong to a different planet population than \planetname{} and thus may be affected by additional effects such as photo-evaporation. In contrast to Fig. \ref{fig::subjov2}, Fig. \ref{fig::subjov1} does not show a significant change in the shape of the planetary population of hot Jupiters when the stellar parameters are taken into account.\\

Together with inflation and high-eccentricity migration, photo-evaporation might also play a role in shaping the sub-Jovian desert, especially at the lower boundary. Photo-evaporation is caused by the EUV or X-ray radiation of its hosts star. 
The planet is mainly exposed to this radiation in its early evolutionary stages, and for most planets, we have no knowledge of the amount of EUV or X-ray radiation that the planet received.
As a first estimate, we suggest that the equilibrium temperature is a better approximation for the amount of EUV or X-ray heating or radiation than the orbital period. 

\begin{figure}
        \centering
        \includegraphics[width=0.95\columnwidth]{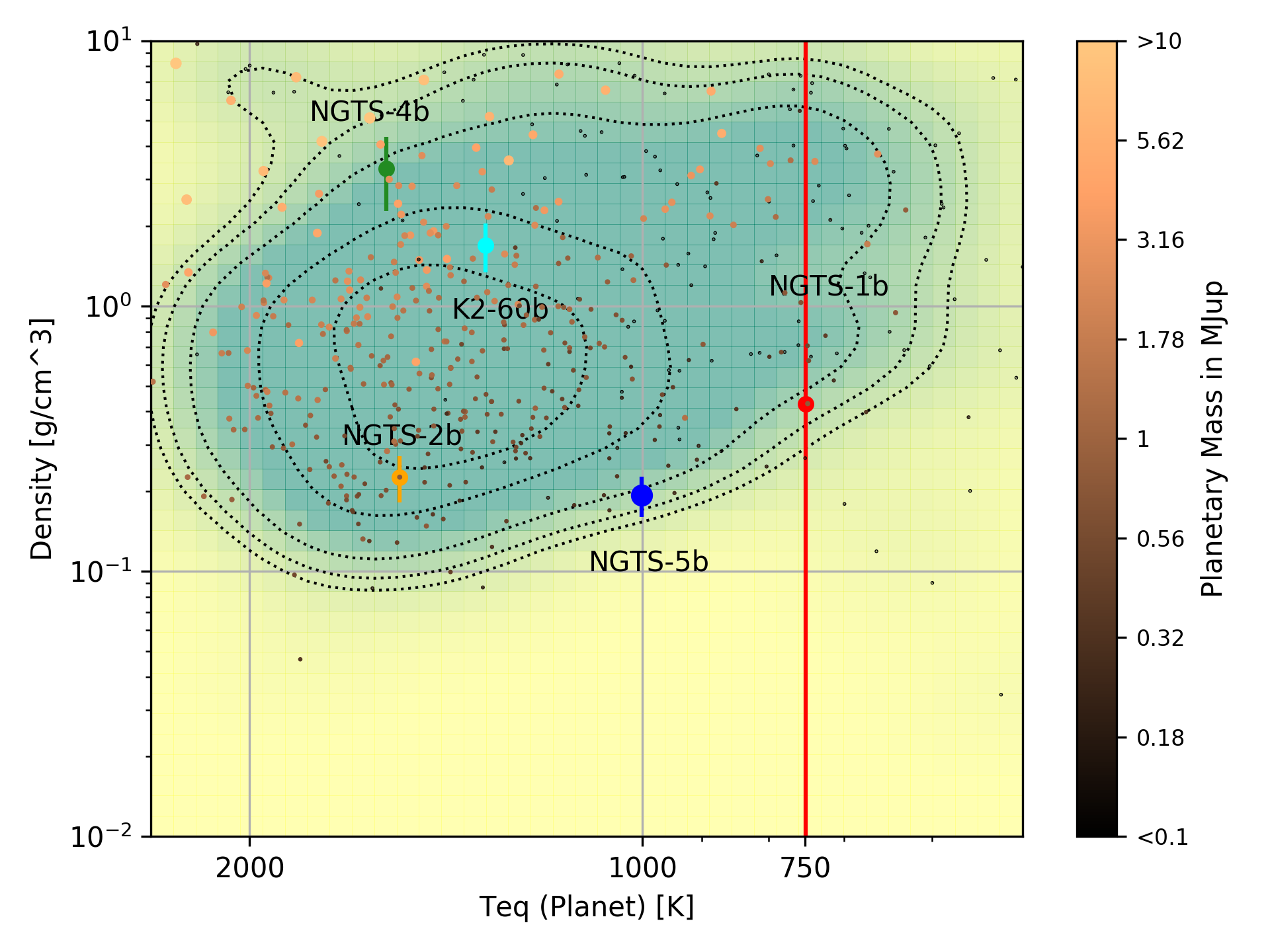}
    \caption{Density over equilibrium temperature colour-coded by planetary mass. Data and background are the same as above.}
    \label{fig::dens}
\end{figure}

In addition to its position in regard to the sub-Jovian desert, \planetname{} is also interesting because of its unusually low density. In Figure \ref{fig::dens} we show the density of planets that have been reported in the literature as a function of the equilibrium temperature. \planetname{} is at the lower limit of what has been found so far in its temperature regime. 
The lower boundary of the planetary population in this parameter space might be directly related to inflation. \planetname{} lies directly at the lower limit of observed densities in the temperature regime around $T_{eq}\approx 1000K$.
The contour lines for the planet population point density show a sharp boundary for the lowest possible densities depending on the equilibrium temperature. 

\section{Conclusions}\label{sec::con}
We reported the discovery of \planetname{} with  a mass of \SecondaryMass{} \Mjup \, and radius of \SecondaryRadius{} \Rjup.
Its low mass places \planetname{} at the edge of the sub-Jovian desert. \starname{} is expected to be observed by TESS in sector 24. The TESS sector 24 observations are scheduled to start in April 2020 and continue into May 2020.\\

A comparison of the upper boundary of the sub-Jovian desert as defined by the point density contour lines in Figs. \ref{fig::subjov1} and \ref{fig::subjov2} to the sub-Jovian desert as empirically determined by \citet{Mazeh2016} reveals discrepancies in the slope. The theoretical upper boundary that is due to high-eccentricity migration as given in \citet{Owen2018} agrees better with the observed point density.   

To allow us to account for the stellar type when we set this planet in context with the known planet population, we introduced a presentation of the sub-Jovian desert in a parameter space that includes the influence of the host star in the planetary evolution. Using the mass ratio between the planet and host star instead of the planetary mass, we accounted for the stellar mass that affects the high-eccentricity migration. To take the luminosity of the host star into account when we studied the inflation of hot Jupiters, we considered the planetary radius as a function of equilibrium temperature instead of the orbital period. \\
Taking the host star and its mass and luminosity into account changed the location of \planetname{} and other planets especially around low-mass stars in regard to the sub-Jovian desert. \planetname{} is now at the boundary of the sub-Jovian desert in regard to its mass, when otherwise it would lie more in the centre of the desert. Judging from its radius, it is clearly outside the desert and at the upper edge of a population of inflated planets. Together with the low mass, this leads to a very low density for this equilibrium temperature. The density as a function of equilibrium temperature places \planetname at the lower edge of observed densities. The observed significant changes in the position of single objects in regard to the sub-Jovian desert, or more generally, in regard to planet populations depending on the stellar type of the host star, highlights the care that is required when planet populations are to be discussed. We showed that in case of the sub-Jovian desert, a population analysis that does not take the host star into account might lead to incorrect findings. In order to simplify a concept such as the sub-Jovian desert, which might depend on several parameters, into a 2D plot, the parameter space needs to be selected carefully. We suggest that the sub-Jovian desert might not be displayed not by planetary mass or radius over orbital period, as shown in Figs. \ref{fig::subjov_mazeh1} and \ref{fig::subjov_mazeh2}, but by taking the stellar type into account, as shown in Figs. \ref{fig::subjov1} and \ref{fig::subjov2}.\\

Our presentation of the sub-Jovian desert, which includes information about the host star, will allow us to directly compare observed planet populations of a inhomogeneous sample with model predictions. With red-sensitive planet surveys such as TESS and NGTS, we will continue to detect planets around late-type stars. For these planets it is important to take the host star into account when the planet population in general and the sub-Jovian desert specifically are studied.

\begin{acknowledgements}
This work is based on data collected under the NGTS project at the
ESO La Silla Paranal Observatory. The NGTS facility is
operated by the consortium institutes with support from
the UK Science and Technology Facilities Council (STFC)
project ST/M001962/1. This paper uses observations made
at the South African Astronomical Observatory (SAAO).
The contributions at the University of Warwick by PJW,
RGW, DLP, DJA, BTG and TL have been supported
by STFC through consolidated grants ST/L000733/1 and
ST/P000495/1. Contributions at the University of Geneva
by DB, FB, BC, LM, and SU were carried out within the
framework of the National Centre for Competence in Research "PlanetS" supported by the Swiss National Science Foundation (SNSF). The contributions at the University
of Leicester by MRG and MRB have been supported by
STFC through consolidated grant ST/N000757/1. CAW acknowledges support from the STFC grant ST/P000312/1.
EG gratefully acknowledges support from Winton Philanthropies in the form of a Winton Exoplanet Fellowship.
MNG is supported by the STFC award reference 1490409 as well as the Isaac Newton Studentship.
JSJ acknowledges support by Fondecyt grant 1161218 and
partial support by CATA-Basal (PB06, CONICYT). DJA
gratefully acknowledges support from the STFC via an
Ernest Rutherford Fellowship (ST/R00384X/1). PE, ACh, and HR
acknowledge the support of the DFG priority program SPP
1992 "Exploring the Diversity of Extrasolar Planets" (RA
714/13-1). LD acknowledges support from the Gruber Foundation Fellowship. The research leading to these results has received funding from the European Research Council under the FP/2007-2013 ERC Grant Agreement number 336480 and from the ARC grant for Concerted Research
Actions, financed by the Wallonia-Brussels Federation. This
work was also partially supported by a grant from the Simons Foundation (PI Queloz, ID 327127).  This work has made use of data from the European Space Agency
(ESA) mission Gaia (https://www.cosmos.esa.int/gaia),
processed by the Gaia Data Processing and Analysis Consortium (DPAC, https://www.cosmos.esa.int/web/gaia/dpac/consortium). Funding for the DPAC has been provided by national institutions, in particular the institutions participating in the Gaia Multilateral Agreement.
PyRAF is a product of the Space Telescope Science Institute, which is operated by AURA for NASA.
This research has made use of the NASA Exoplanet Archive, which is operated by the California Institute of Technology, under contract with the National Aeronautics and Space Administration under the Exoplanet Exploration Program.
\end{acknowledgements}

%-------------------------------------------------------------------

\bibliographystyle{aa}
\bibliography{ngts3}

\end{document}